%% file: 0_hw-attest.tex
\documentclass{sig-alternate} 
\newif\ifsubmission
\submissiontrue 
\submissionfalse 

\newif\ifshepherding
\shepherdingfalse %

\newif\ifanonymous
\anonymousfalse

\newif\iffullversion
\fullversionfalse

\usepackage{tikz}
\usepackage{longtable}
\usepackage{booktabs}
\usepackage{graphicx}
\usepackage{epstopdf}
\usepackage{epsfig}
\usepackage{amsmath}
\usepackage{multirow}
\usepackage{amsmath}
\usepackage{amsfonts}
\usepackage{amssymb}
\usepackage{algorithmic}
\usepackage{algorithm}
\usepackage{tikz}
\usepackage{tikzscale}
\usepackage{mathtools}
\usepackage{placeins}
\usepackage{balance}
\usepackage{paralist}
\usepackage{pifont}
\usepackage{relsize}
\let\llncssubparagraph\subparagraph
\let\subparagraph\paragraph
\usepackage[compact]{titlesec}
\let\subparagraph\llncssubparagraph

\PassOptionsToPackage{bookmarks=false}{hyperref}

\usepackage{ellipsis,ragged2e,marginnote}
\usepackage{enumitem}
\usepackage{listings}
\usepackage{array}
\setlist[itemize]{noitemsep,nolistsep}
\usepackage[tracking=true]{microtype}
\DeclareMicrotypeSet*[tracking]{my}{ font = */*/*/sc/* }
\SetTracking{ encoding = *, shape = sc }{ 45 }

\usepackage{float} 
	\floatstyle{ruled}

\usepackage{color}
\definecolor{darkred}{rgb}{.6,0,0}
\definecolor{darkgreen}{rgb}{0,.4,0}
\definecolor{darkblue}{rgb}{0,0,.6}

\usepackage[colorlinks=true,
	citecolor=darkgreen,
	urlcolor=darkblue,
	linkcolor=darkred
]{hyperref}

\usepackage{xspace}

\usepackage{footnote}

\makesavenoteenv{tabular}

\usepackage{numprint}
\npthousandsep{\,}\npthousandthpartsep{}\npdecimalsign{.}

\newcolumntype{x}[1]{
>{\centering\hspace{0pt}}p{#1}}%

\input{macros}

\usetikzlibrary{arrows,decorations.pathmorphing,backgrounds,fit,positioning,shapes.symbols,chains}

\makeatletter
\def\@copyrightspace{\relax}
\makeatother
\def\sharedaffiliation{%
\end{tabular}
\begin{tabular}{c}}

\begin{document}


\title{\tool: \underline{L}ow-\underline{O}verhead Control \underline{F}low \underline{AT}testation in Hardware} 
\ifanonymous
\else
\numberofauthors{1}
\author{
\alignauthor
Ghada Dessouky\textsuperscript{$^1$},
Shaza Zeitouni\textsuperscript{$^1$},
Thomas Nyman\textsuperscript{$^{2,3}$},
Andrew Paverd\textsuperscript{$^2$},
Lucas Davi\textsuperscript{$^4$},
Patrick Koeberl\textsuperscript{$^5$},
N. Asokan\textsuperscript{$^2$},
Ahmad-Reza Sadeghi\textsuperscript{$^1$}\\
\sharedaffiliation
      \affaddr{\textsuperscript{$^1$}Technische Universit\"at Darmstadt, Germany}\\
      \affaddr{\{ghada.dessouky,shaza.zeitouni,ahmad.sadeghi\}@trust.tu-darmstadt.de}\\
\sharedaffiliation
      \affaddr{\textsuperscript{$^2$}Aalto University, Finland}\\
      \affaddr{thomas.nyman@aalto.fi, andrew.paverd@ieee.org, asokan@acm.org}\\
\sharedaffiliation
      \affaddr{\textsuperscript{$^3$}Trustonic, Finland}\\
      \affaddr{thomas.nyman@trustonic.com}\\
\sharedaffiliation
      \affaddr{\textsuperscript{$^4$}University of Duisburg-Essen, Germany}\\
      \affaddr{lucas.davi@uni-due.de}\\
\sharedaffiliation
      \affaddr{\textsuperscript{$^5$}Intel Labs, Germany}\\
      \affaddr{patrick.koeberl@intel.com}\\
}
\fi

\maketitle \label{r:title}

\begin{abstract}
  \input{abstract}

\end{abstract}

\input{1_intro}
\input{2_problem_setting} 
\input{3_system_model} 
\input{4_design} 
\input{5_impl} 
\input{6_eval} 
\input{7_related}
\input{8_conc}

\bibliographystyle{abbrv}
\bibliography{hw-attest}


\end{document}

%% file: macros.tex
\ifsubmission

  \newcommand{\tsc}[1]{}
  \newcommand{\asa}[1]{}
\else

  \newcommand{\tsc}[1]{\textcolor{red}{\footnote{\textcolor{red}{tsc: #1}}}}
  \newcommand{\asa}[1]{\textcolor{red}{\footnote{\textcolor{red}{asa: #1}}}}
\fi

\let\orgautoref\autoref
\renewcommand{\autoref}
        {\def\equationautorefname{Equation}%
         \def\figureautorefname{Figure}%
         \def\subfigureautorefname{Figure}%
         \def\Itemautorefname{Item}%
         \def\tableautorefname{Table}%
         \def\algorithmautorefname{Figure}%
         \def\paragraphautorefname{Paragraph}%
         \def\sectionautorefname{Section}%
         \def\subsectionautorefname{Section}%
         \def\subsubsectionautorefname{Section}%
         \def\chapterautorefname{Chapter}%
         \def\partautorefname{Part}%
         \def\goalautorefname{Goal}%
         \def\reqautorefname{Req.}%
         \def\adviceautorefname{Rule}%
         \def\parameterautorefname{Param.}%
         \def\definitionautorefname{Definition}%
         \def\theoremautorefname{Theorem}%
         \orgautoref}

\newcommand{\remove}[1]{}

\newcommand{\sect}[1]{\S\ref{#1}}

\newcommand{\myparagraph}[1]{\vspace{4pt}\par{\noindent\bf#1.}}
\renewcommand{\paragraph}{\myparagraph}

\newcommand{\prv}{\ensuremath {\mathcal P}\xspace}
\newcommand{\vrf}{\ensuremath {\mathcal V}\xspace}
\newcommand{\auth}{\ensuremath {A }\xspace}
\newcommand{\meta}{\ensuremath {L }\xspace}

\newcommand{\stp}{\ensuremath {\mathcal (Src, Dest)}\xspace}
\newcommand{\dOne}{\ding{182}}
\newcommand{\dTwo}{\ding{183}}
\newcommand{\dThree}{\ding{184}}

\newcommand{\tool}{LO-FAT\xspace}

%% file: abstract.tex
Attacks targeting software on embedded systems are becoming increasingly prevalent. 
Remote attestation is a mechanism that allows establishing trust in embedded devices.
However, existing attestation schemes are either static and cannot detect control-flow attacks, or require instrumentation of software incurring high performance overheads.
To overcome these limitations, we present \tool{}, the first \emph{practical hardware-based} approach to control-flow attestation.
By leveraging existing processor hardware features and commonly-used IP blocks, our approach enables efficient control-flow attestation without requiring software instrumentation.
We show that our proof-of-concept implementation based on a RISC-V SoC incurs no processor stalls and requires reasonable area overhead.

%% file: 1_intro.tex
\section{Introduction}\label{sect:intro}

Embedded systems have been facing a variety of security challenges for decades~\cite{viega-sp-2012} which are becoming increasingly prevalent with emerging trends such as collaborative Internet of Things (IoT). A recent prominent example is \emph{Mirai} malware\footnote{\url{https://www.incapsula.com/blog/malware-analysis-mirai-ddos-botnet.html}} in October 2016, where a series of Distributed Denial-of-Service (DDoS) attacks against the DNS system disrupted a number of prominent websites.
These attacks were perpetrated by IoT devices, including routers, DVRs, and web-enabled security cameras, that had been compromised by the Mirai malware.

Increasingly, attacks against embedded systems aim to exploit software vulnerabilities. In 2015, a remotely exploitable buffer overflow vulnerability was found in the \emph{USB over IP} software used in millions of residential gateways and wireless routers supplied by prominent manufacturers\footnote{\url{http://blog.sec-consult.com/2015/05/kcodes-netusb-how-small-taiwanese.html}}.
In 2014, a memory corruption flaw was found in the embedded webserver software used by over 200 different models of embedded devices, affecting at least 12 million devices, many of which still remain vulnerable today\footnote{\url{http://mis.fortunecook.ie/}}.

\emph{Remote attestation} is an important class of security mechanisms designed to detect software attacks.
In principle, remote attestation allows one entity (the \emph{verifier}) to ascertain the precise state of the software running on a remote system (the \emph{prover}).
However, most attestation schemes are \emph{static} in that they attest the software initially loaded by the prover before it begins executing.
Although useful, this still leaves the system vulnerable to \emph{run-time} software attacks. If the adversary gains control of the stack or heap, (s)he can alter control-flow information to subvert the control flow of the target program, and mount a \emph{code-reuse attack}.  
Similarly, in \emph{non-control data} attacks~\cite{non-control-data}, the adversary modifies strategic data variables to cause a permissible but unintended control flow change (e.g., executing a privileged instruction sequence). 
Traditionally, code-reuse attacks are mitigated using techniques such as control-flow integrity (CFI)~\cite{Abadi09}. 
However, CFI cannot prevent non-control data attacks, since these do not violate control-flow integrity.
Neither of these types of attacks can be detected by static attestation.

To overcome these challenges \emph{control-flow attestation}~\cite{Abera2016a} was proposed very recently, enabling the prover to precisely report the control flow of application software to the verifier while giving assurance on control-flow integrity and detection of non-control data attacks. The attestation mechanism of ~\cite{Abera2016a} requires an isolated execution environment (e.g., ARM TrustZone, Intel SGX) to protect it against potentially compromised application software.
However, implementing control-flow attestation in software has two limitations: Firstly, in order to detect control-flow events, the application software must be \emph{instrumented} prior to deployment. Non-instrumented or incorrectly-instrumented software cannot be attested. The instrumentation rewrites all control-flow instructions (e.g., \texttt{branch}, \texttt{return}, etc.) in order to transfer control to the attestation software. Secondly, the attestation software runs on the main processor which incurs significant performance penalties because single control-flow instructions are essentially replaced with relatively many numbers of instructions in order to track and record the control-flow event (e.g., update a running hash value).
%
As we elaborate in \sect{sect:related}, some existing hardware approaches, such as debugging and tracing features in modern processors~\cite{Soffa11,Intel-IPT} or hardware security architectures~\cite{basakdac2016,karri2015,Sofia}, can be used to record control flow information.
However, due to the overhead they incur or the type of information they record, these approaches are not well-suited for control-flow \emph{attestation}.

\textbf{Goals and Contributions.}
To overcome the limitations of a software solution, we introduce a practical hardware-based \underline{L}ow-\underline{O}verhead Control \underline{F}low \underline{AT}testation architecture, \tool{}.
Unlike software implementations, \tool{} can handle \emph{unmodified application software} without instrumentation, meaning that it is transparent to legacy software.
By recording the control flow in hardware in parallel to the main processor, \tool{} does not stall the application software, thus eliminating the performance overhead of attestation in software.
\tool{} leverages existing processor features and commonly-used IP blocks and can feasibly be implemented on typical embedded systems hardware platforms.


The main contributions of this paper are:

\begin{compactitem}

\item Design of \tool, a hardware-based scheme for control-flow attestation, providing the \emph{same security guarantees} as previous software schemes, without the performance overhead or the need for software instrumentation (\sect{sect:design}).

\item An integrated optimization for eliminating redundant attestation computation (e.g., avoiding duplication when attesting loops) and reducing the burden on the verifier (\sect{sect:design}).

\item A proof-of-concept implementation of \tool\ on the new open-source RISC-V architecture targeting the Pulpino core for single-threaded embedded system software (\sect{sect:impl}).

\item A systematic evaluation of \tool{} in terms of the required hardware area and performance benefits (\sect{sect:eval}).

\end{compactitem}

%% file: 2_problem_setting.tex
\section{Problem Setting and Challenges}\label{sect:problem}
 
\begin{figure} [h]
\centering
\includegraphics[width=0.75\columnwidth]{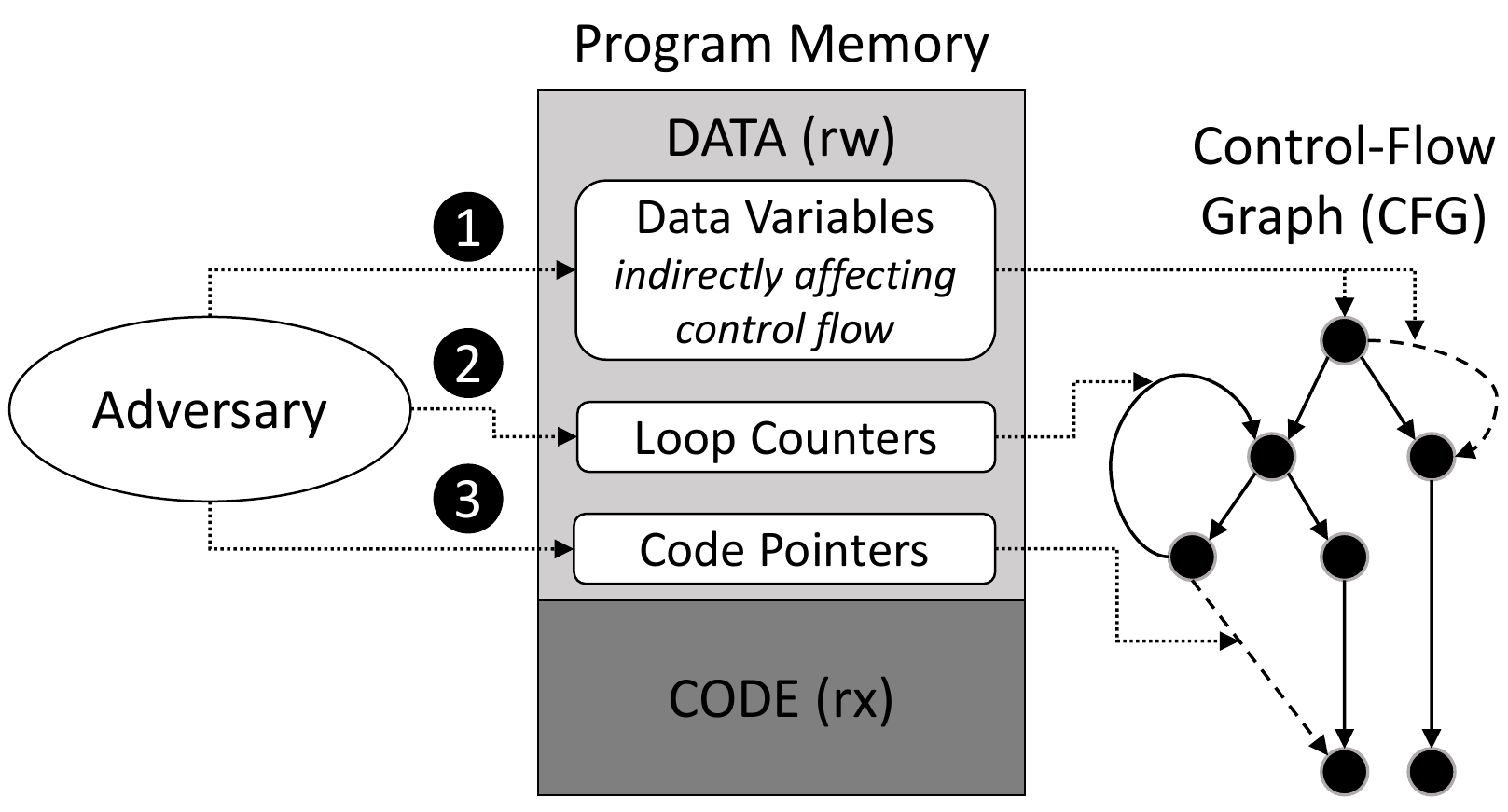}
\caption{Overview on run-time attack classes}
\label{fig:runtime-attacks}
\end{figure}

Remote attestation provides a well-known mechanism for detecting malware on a device. However, existing conventional (binary) attestation cannot detect run-time exploitation techniques, since run-time attacks do not not modify the program binary. Such attacks aim to subvert the intended control flow of the targeted program while it is executing. An overview of different classes of such attacks is shown in \autoref{fig:runtime-attacks}. In general, a program reserves dedicated  memories for data and code. The former is marked as readable and writable~\emph{(rw)}, whereas the latter is as readable and executable~\emph{(rx)}. This ensures that code cannot be executed from data memory, and code memory cannot be overwritten. Furthermore, any program can be abstracted through its corresponding control-flow graph (CFG) that encapsulates the valid paths a program should follow at run-time. 

We can distinguish three classes of run-time attacks: \dOne\ non-control-data attacks that indirectly affect the control flow of a program, \dTwo\ corruption of loop counter variables, and \dThree\ code-pointer overwrites. The most prominent run-time attacks exploit code-pointer overwrites, i.e., corruption of return addresses and function pointers. For instance, code-reuse attacks such as \emph{Return-oriented Programming} (ROP)~\cite{Shacham07} exploit memory corruption vulnerabilities (e.g., buffer overflows) in the program and then stitch together a malicious sequence of machine code instructions from benign \emph{gadgets} of code already residing in the vulnerable program memory. This is exemplified by a malicious CFG edge (see dashed line for code-pointer overwrite in Figure~\ref{fig:runtime-attacks}). These attacks have been shown to be a realistic threat on many processor architectures, such as Intel x86~\cite{Shacham07}, ARM~\cite{Le11} and embedded systems building on Atmel AVR~\cite{Francillon08}. Although countermeasures against this class of attacks exist, e.g., control-flow Integrity (CFI)~\cite{Abadi09} and code-pointer integrity (CPI)~\cite{Kuznetsov14}, they do not prevent attacks \dOne\ and \dTwo. The so-called \emph{non-control data attacks}~\cite{non-control-data} do not compromise the control flow of a program, but cause unexpected malicious control-flow paths by corrupting data variables. In \dOne, the attacker compromises data variables that are used for security decisions during program execution, e.g., corrupting an authentication variable to execute a privileged but existing path. Attack class \dTwo\ is even more subtle as it only affects the number of times a program loop is executed. This can have severe consequences in the context of embedded system software, e.g., a syringe pump dispenses more liquid than requested (see~\cite{Abera2016a}).

Control-flow attestation can cover these cases by assuring the verifier of the precise run-time control flow of the program on the embedded device. In ~\cite{Abera2016a}, the first control-flow attestation scheme was proposed and implemented. However, it suffers from practical limitations, such as high performance overhead and the need for tedious software instrumentation.

Our work tackles the challenge of detecting attack classes \dOne - \dThree, while addressing the limitations of recently proposed software-based control-flow attestation~\cite{Abera2016a} by presenting \tool, an efficient hardware-only solution.

%% file: 3_system_model.tex
\section{System Model}\label{sect:model}

\begin{figure}[t]
\centering
\includegraphics[width=\columnwidth]{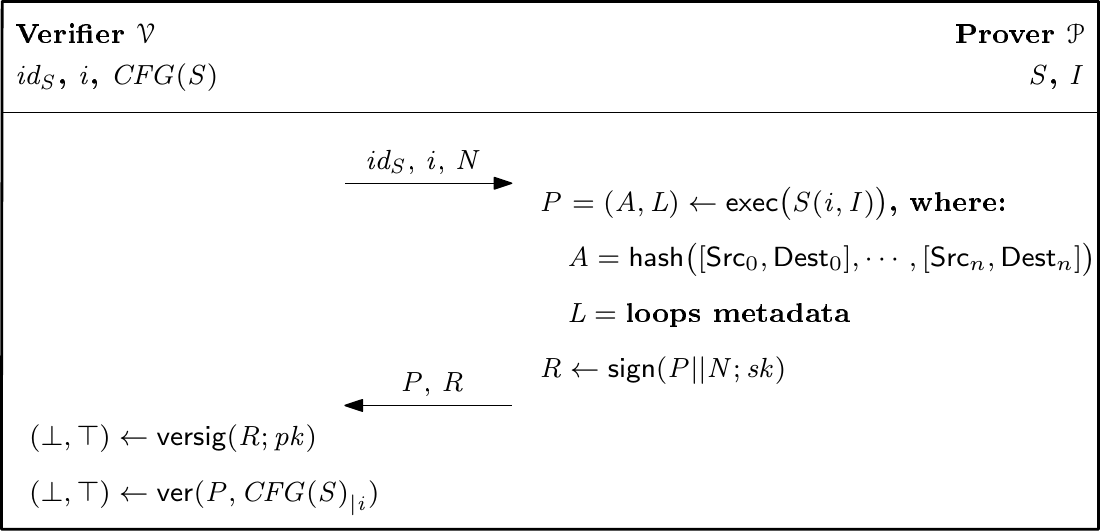}
\caption{Attestation protocol of \tool}
\label{fig:model}
\end{figure}

\autoref{fig:model} depicts the attestation protocol of \tool: the verifier \vrf aims to attest the run-time control-flow (execution path) of the Program $S$ on a remote embedded system -- the prover \prv. We assume that both \vrf and \prv have access to the program $S$ in binary form and that conventional static (binary) attestation assures \prv is executing the correct and unmodified program $S$. 

First, \vrf performs a one-time offline pre-processing step to generate the CFG of $S$ (including expected loop execution information) by means of static or dynamic analysis. Next, \vrf initiates the protocol by sending \prv the program input $i$ for the program ID $id_S$, and the nonce $N$ to ensure freshness of the attestation response. \prv executes $S$ with verifier input $i$ and a set of malicious adversary inputs $I$. In fact, the untrusted inputs received may corrupt the control-flow by means of the attack techniques described in ~\sect{sect:problem}. While $S$ executes, \tool captures the control-flow transitions and generates a cumulative authenticator \auth of the control-flow path taking source and destination address \stp of each branch as input. Naively storing and transmitting every single executed instruction to \vrf would incur impractical memory, power and communication overheads, especially for resource-constrained embedded devices. Hence, \tool\ follows the idea outlined in~\cite{Abera2016a} and computes a cumulative cryptographic hash of the executed path. In addition, it also produces auxiliary metadata \meta to track program loop paths and their number of iterations (including recursive functions) thereby covering attacks of class \dTwo\ in ~\autoref{fig:runtime-attacks}. Together \auth and \meta form a unique program path $P$. Lastly, upon program exit, \prv generates the \emph{attestation report} $R=sign(P||N;sk)$, under the signing key $\mathit{sk}$, which is stored by \prv in hardware-protected secure memory, e.g., a register that is accessible only to \tool. Upon receiving $R$, \vrf verifies the signature using the verification key $pk$. Next, \vrf checks whether the reported path $P$ resembles a valid path in CFG under input $i$. If true, \vrf is assured of \prv's execution.

\noindent\textbf{Adversary Model and Assumptions.} 
We assume a strong adversary that has full control over the \emph{data memory} of \prv and can utilize standard memory corruption vulnerabilities to modify arbitrary writable memory locations. However, the adversary cannot modify program code at run-time (marked as \emph{rx}) and cannot modify memory used by \tool\ itself (due to hardware protection). Note that similar to all attestation schemes we consider software-only attacks and hence physical attacks on \prv's device are out of scope in this work. 
Also note that our scheme can detect attacks that affect the program's control-flow, but not pure data-driven attacks (that do not affect any control-flow) such as \emph{data-oriented programming} attacks, which remain an open research problem~\cite{dop}. 

%% file: 4_design.tex
\section{\tool Design}\label{sect:design}

%

\begin{figure}[ht]
\includegraphics[width=0.50\textwidth]{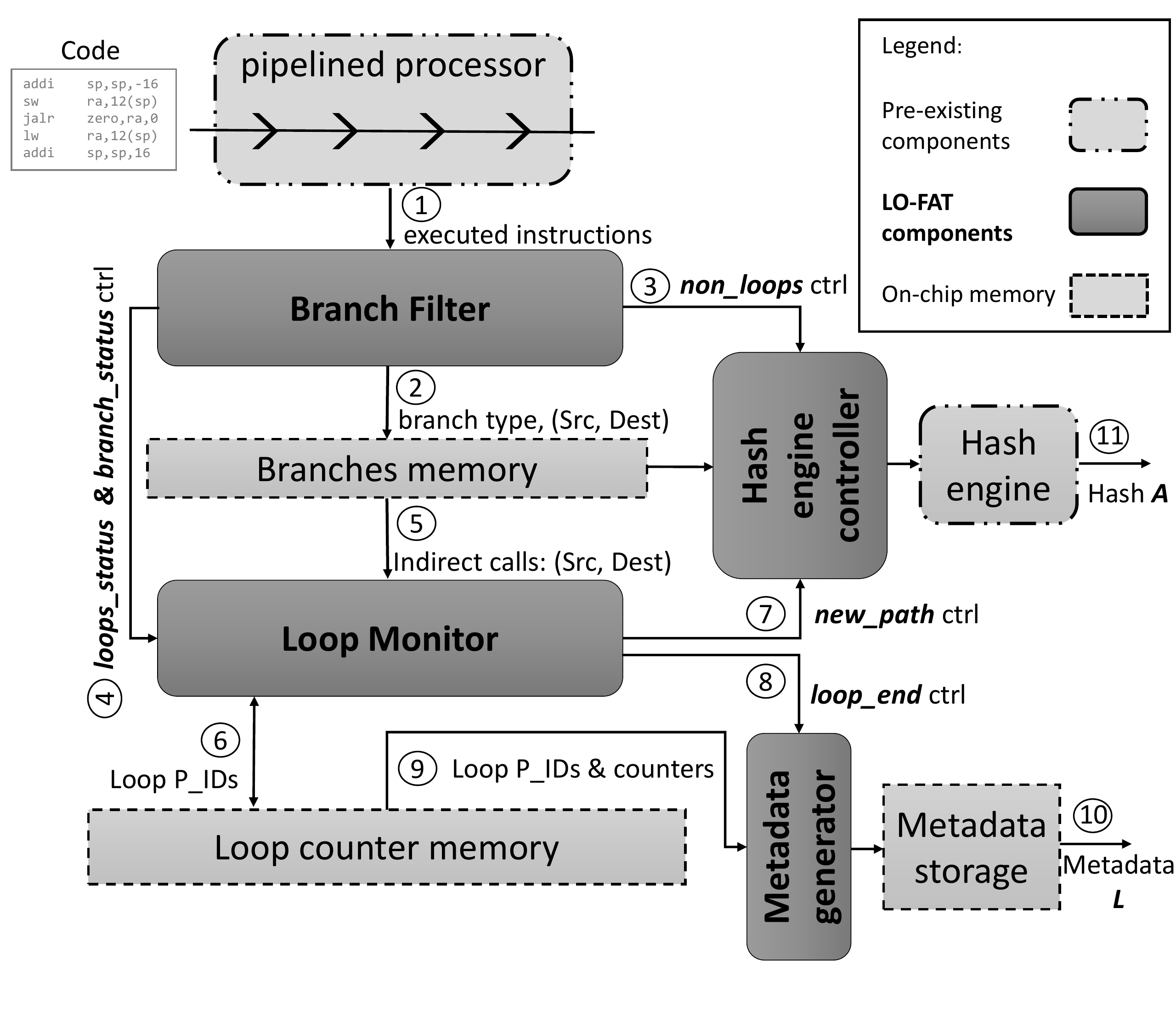}
\caption{Architecture of \tool.}
\label{fig:arch}
\end{figure}

\autoref{fig:arch} illustrates our architecture for \tool and how it interfaces with the processor pipeline. The proposed scheme exploits branch tracking functionality inherent in any processor pipeline and re-usable IP cores such as the hash engine. We extend these with additional logic to achieve efficient tracing of control-flow information. The main \tool components are the branch filter and the loop monitor. The former extracts branch instructions from the processor as it executes the attested code segment while the latter monitors program loops.

\noindent\textbf{Branch Filter.}
Upon code execution, the branch filter, which is tightly coupled to the processor, extracts the current program counter and instruction executed per clock cycle. Then it filters in every \texttt{branch}, \texttt{jump} and \texttt{return} instruction since these are the relevant instructions for control-flow attestation.
The branch filter outputs a concise representation of every executed branch instruction with its source and destination address pair \stp into a dedicated branches memory and detects whether the intercepted branch is within a program loop. If not, the branch filter enables hashing of \stp. Branches inside a program loop require special treatment in \tool, because (i)~loop counter manipulation may compromise the program's control-flow in a malicious way (\sect{sect:problem}), and (ii)~naively hashing each loop iteration and path leads to a combinatorial explosion of valid hash values~\cite{Abera2016a}. As such, we design \tool to compress control-flow information associated with loops efficiently. As mentioned earlier in \sect{sect:model}, we report each loop path and its number of iterations as auxiliary metadata \meta. However, doing so in hardware is challenging, i.e., in contrast to the most related work C-FLAT, since we do not use code instrumentation to preserve legacy compliance. Hence, the branch filter must detect and identify loop entry and exit points and their depth at run-time without instrumentation aid. We describe in \sect{sect:loop_detect} how we tackled this challenge.

\noindent\textbf{Loop Monitor.}
When a loop is encountered, the branch filter forwards the loop entry and exit to the loop monitor. The loop monitor identifies and tracks program loops (including nested loops). When a branch inside a program loop is encountered, the branch filter forwards this information to the loop monitor which in turn encodes each path inside the loop uniquely. Simultaneously, \stp of each \texttt{branch} remains stored in the branches memory. 

Another major challenge concerning loops is the hash computation and attestation overhead incurred by hashing each loop iteration. In \tool, we significantly reduce the hash computation cost by only hashing each loop path once and keeping an iteration counter for each unique loop path. 
To achieve this, \tool\ generates a unique path encoding for each loop path and associates an on-chip loop counter with it. The loop monitor indicates newly observed loop paths to the hash engine controller in order to hash its corresponding \stp from the branches memory. On the other hand, once the same loop path executes, \tool\ only needs to increment the counter, i.e., not requiring further hash operations. 

Upon loop exit, the loop monitor requests the metadata generator to assemble the loop auxiliary metadata based on the loops memory which contains the unique loop path encodings, their number of iterations, and indirect branch targets. This information is stored on-chip and is appended to the final hash value \auth computed at the end of the attested execution. Finally, a digital signature $R$ is computed over the hash value \auth, metadata \meta and nonce $N$ and sent to \vrf for attestation (as per our protocol outlined in \sect{sect:model}).

%% file: 5_impl.tex
\section{Implementation}\label{sect:impl}

\subsection{Loop Handling} \label{sect:loop_detect}  
\label{sect:loop_track}
\textbf{Detecting loops.}
As shown in \autoref{fig:arch}, the branch filter unit traces the instruction (and its address) executed per clock cycle and filters in {\larger\textcircled{\smaller[2]1}} every \texttt{branch}, \texttt{jump} and \texttt{return} instruction.
It outputs a concise representation of every executed branch instruction with its \stp-pair into a dedicated branch buffer ({\larger\textcircled{\smaller[2]2}}). To compress the control-flow trace for loops, the branch filter has to detect loops. If the intercepted branch is not in a loop, the branch filter sends the control signal \emph{non\_loops\_ctrl} to the existing hash engine controller to compute a hash over \stp in {\larger\textcircled{\smaller[2]3}}. Otherwise, the branch filter forwards the loop status (entry and exit) to the loop monitor and its depth (in case of nested loops) via the \emph{loops\_status\_ctrl} signals ({\larger\textcircled{\smaller[2]4}}).

To enable efficient run-time loop detection, we utilize a property of RISC architectures that implement a \emph{link-register}, such as PowerPC, ARM, SPARC, and RISC-V. \tool uses a simple heuristic to differentiate between backward branches that constitute loops, and branches for subroutine calls where the call target resides earlier in memory. Since subroutine calls use instructions that update the \emph{link-register}, we consider the target of each \emph{non-linking} backwards branch as a \emph{loop entry node}. The basic block proceeding the branch instruction is considered a \emph{loop exit node}. We base our heuristic on our observations of the RISC-V compiler assembly and the calling convention described in the instruction manual: any subroutine call with multiple call sites must be \emph{linking} and updates the \emph{link-register}. Subroutines with a single call site are still compiled as a \emph{linking} branch or are optimized by traditional inlining using the RISC-V compiler.

The addresses of the entry and exit nodes of each loop are stored in registers by the loop detector and used to detect and track loop iterations and loop depth at run-time when executing nested loops. The number of loop iterations is determined by recording the number of times the loop entry node is entered within the loop. Loop termination is detected by tracking if execution proceeds to or past the currently active loop exit node, either as the result of sequential execution (e.g. in the case of a conditional branch) or a non-linking branch (e.g. break). Loop execution status is forwarded using the \emph{loops\_status\_ctrl} signals to the loop monitor, as shown in ~\autoref{fig:arch}.

\begin{figure}[ht]
\centering
\includegraphics[width=\columnwidth]{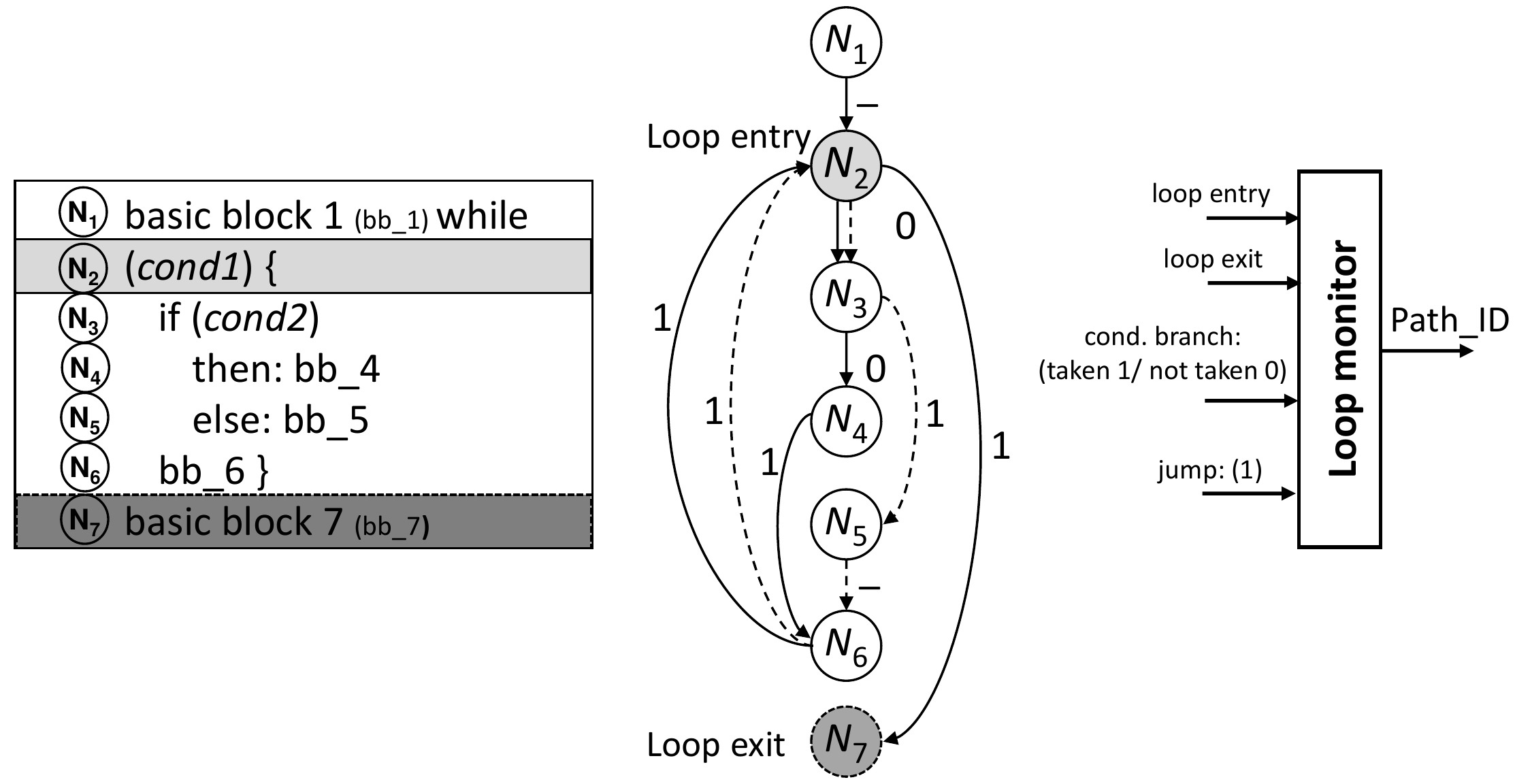}
\caption{CFG for pseudo-code and its layout of instructions in memory.}
\label{fig:loop_cfg}
\end{figure}

\textbf{Tracking loops.}
As shown in \autoref{fig:arch}, the loop monitor receives \emph{branch\_status\_ctrl} signals from the branch filter to describe the type of intercepted branch instruction and its \stp ({\larger\textcircled{\smaller[2]5}}). This branch tracking mechanism allows the loop path encoder to uniquely encode paths as they occur. Simultaneously, \stp\ of each \texttt{branch} along the executing loop path remain stored in the branches memory. 

\autoref{fig:loop_cfg} shows a sample pseudo-code and its CFG according to how the instructions would be laid out in code memory to illustrate how the loop monitor encodes the loop paths. The example code shows a \emph{while-loop} with an \emph{if-else} statement inside. Each basic block in the pseudo-code is represented by a node in the CFG and numbered accordingly, with loop entry and exit nodes also indicated.
Within this simple loop, there are only 2 valid paths: bold path $N_{2} \rightarrow N_{3} \rightarrow N_{4} \rightarrow N_{6} \rightarrow N_{2}$ and dashed path $N_{2} \rightarrow N_{3} \rightarrow N_{5} \rightarrow N_{6} \rightarrow N_{2}$.

For every \texttt{conditional branch}, the processor evaluates the condition and either jumps to the computed target address (branch is taken), or continues sequentially to the next instruction address in memory (branch is not taken). Processors commonly track this branching behavior in the pipeline and may encode a taken/not-taken branch with '1'/'0'. 
This branch information is extracted from the processor by the branch filter and used by the loop monitor to uniquely identify and encode paths within each loop with a unique path\_ID, as shown in \autoref{fig:loop_cfg}. In \autoref{fig:loop_cfg}, the dashed path $N_{2} \rightarrow N_{3} \rightarrow N_{5} \rightarrow N_{6} \rightarrow N_{2}$ is encoded as `011' and bold path $N_{2} \rightarrow N_{3} \rightarrow N_{4} \rightarrow N_{6} \rightarrow N_{2}$ as `0011'. Other path encodings are considered invalid and detected by the \vrf.

Once a loop path is completed, this unique path\_ID is used to index \emph{loop counter} memory, in which the number of iterations for each corresponding path is saved ({\larger\textcircled{\smaller[2]6}}) in \autoref{fig:arch}. A counter value of zero indicates the first time a particular path is executed. This is forwarded by the loop monitor into the hash engine controller using \emph{new\_path\_ctrl} signals ({\larger\textcircled{\smaller[2]7}}) to enable hashing of corresponding \stp pairs. Otherwise, the counter is simply incremented.

To ensure constant-time, single-cycle memory access latency, we implement \emph{loop counter memory} as on-chip memory indexed by the unique loop path encodings. However, this consumes a dedicated sparsely-utilized memory which is often a constrained resource on low-end embedded devices. In light of this, \tool allows configuring the granularity of the control-flow tracking according to the availability of memory resources.

Once a loop exits, this is identified by the loop monitor and indicated in the \emph{loop\_end\_ctrl} signals sent to the metadata generator ({\larger\textcircled{\smaller[2]8}}). The metadata generator assembles the loop auxiliary metadata from the loops memory - this consists of the unique loop path encodings in order of first occurrence, the number of iterations of each path, and the indirect branch targets encountered in this loop ({\larger\textcircled{\smaller[2]9}}). This fine-grained auxiliary information on loop execution is stored on-chip ({\larger\textcircled{\smaller[2]10}}) and is appended to the final hash value computed at the end of the attested execution ({\larger\textcircled{\smaller[2]11}}). Finally, a digital signature is computed over the hash value, metadata and nonce $N$, and sent to \vrf for attestation. Handling indirect branches in loops is yet another implementation challenge we discuss next.

\subsection{Handling Indirect Branches in Loops}\label{sect:indir_br}
Indirect branches can involve any arbitrary number of targets which can never be exhaustively identified using static analysis. To uniquely identify loop paths with indirect branches (calls and returns), we would need to include the 32-bit target addresses into the path encodings, which would require infeasibly high memory requirements for loop path-indexed memory. Instead, we re-encode the addresses using a smaller number of $n$ bits, allowing a maximum number of ${2}^n$-1 possible targets for each loop. Target addresses are encoded at run-time and stored in a register file, which is implemented as 2 interleaved CAMs to ensure low-latency constant-time access. When a target address is encountered that exceeds the configured limit, we report this in the encoding to the \vrf by an all-zero code. \tool is designed such that the maximum number of branches per loop path and the maximum number of possible target addresses (of indirect branches) to track is configurable in a trade-off between granularity and availability of on-chip memory. Tracking $\ell$ branches per path in a loop requires $8 \times 2^{\ell}$ bits memory.
In our implementation, we configure $n = 4$ to track up to 16 possible indirect branch targets for a given loop and $\ell = 16$ such that \tool can handle a maximum of 16 branches per loop path (every additional indirect branch tracked reduces the maximum number of possible conditional branches by $n$) and depth of up to 3 nested loops, which requires a dedicated 1.5 Mbits memory that is synthesized as block RAM (BRAM) when prototyping on FPGA. Once a loop exists, its memory is re-used for other subsequent loop executions. 

\textbf{Loop metadata.} The measurement in \emph{\auth} is a single hash computation of \stp pairs of executed loop paths. To enable \vrf to reconstruct the final hash value, metadata \emph{\meta} of the loops serves as helper data and provides \vrf with fine-grained insight into the execution of the loops.  \emph{\meta} contains the encodings of executed paths in each loop, the order of first occurrence of each executed path, and number of iterations per loop path and indirect branch targets.

\subsection{Hash Engine}\label{sect:hash}
A single hash measurement \auth is computed on the full execution path, along with auxiliary loop metadata \meta. We employ a SHA-3 512-bit open-source engine\footnote{\url{http://opencores.org/project,sha3}} operating at a maximum clock frequency of 150 MHz. It consists of a permutation module which operates on a message block size of 576-bit. User input is absorbed by the core first into a padding module to assemble the 576-bit block size. Once this padding is full, the permutation module begins computation on input. In \tool, the engine can absorb a 64-bit input \stp-pair every clock cycle into the padding module for 9 clock cycles, after which the 576-bit buffer becomes full and notifies the permutation module to begin its computation. Once full, the padding buffer cannot absorb further input for 3 clock cycles after which it resumes normally. Therefore, a small cache buffer is configured at the hash engine input to prevent dropping of \stp-pairs if they arrive during these cycles where the padding buffer is full. Using this hash engine, an unlimited message size can be hashed while indicating the end of streaming \stp-pairs when the execution of attested software is completed.


%% file: 6_eval.tex
\section{Evaluation}\label{sect:eval}
We present a proof-of-concept implementation of \tool on Pulpino~\cite{pulpino}, the first open-source RISC-V-based microcontroller SoC~\cite{RISCV}. It is based on a single 32-bit 4-stage minimal RISC-V core targeting low-end embedded systems.
We augment the RISC-V processor pipeline to interface with the \tool{} branch filter to extract control-flow signals required for execution flow tracing. \tool{} can be easily integrated into any low-end embedded processor as it does not require modifications to the ISA. 

\subsection{Functionality and Performance}
We integrated \tool with Pulpino and performed cycle-accurate functional simulation of their RTL Verilog source code on ModelSim while Pulpino executed extracted code segments from real embedded applications, such as Open Syringe Pump\footnote{\url{https://hackaday.io/project/1838-open-syringe-pump}}, an open-source open-hardware syringe pump design.
Simulation results confirmed the functionality of \tool in correctly capturing and compressing the control flow (branches, loops, and nested loops) of an uninstrumented application.
Since \tool extracts and filters control-flow events in parallel with the processor, it does not incur any performance overhead for the attested software, as opposed to C-FLAT which incurs attestation overhead that is linearly dependent on the number of control-flow events.
\tool internally incurs latency of 2 clock cycles for branch instructions and loop status tracking and 5 clock cycles at loop exit for completing path\_ID generation and \emph{loop counter} memory access and update. However, \tool simultaneously continues to absorb and process any incoming \stp-pairs to prevent the processor from stalling or dropping trace information. 
Synthesis results using Xilinx Vivado indicate \tool can operate at maximum clock frequency of 80~MHz on a Virtex-7 XC7Z020 FPGA device on a Zedboard. The \tool units are engineered such that they operate on par with Pulpino's clock frequency, while also allowing single-cycle constant-time memory accesses for indirect branches and loops management. Eliminating the CAM access results in a much higher clock frequency if desired.

The length of the auxiliary metadata (\meta{}) that must be sent to \vrf{} depends on the number of loops executed, the number of different paths per loop, and the number of indirect branch targets encountered in the attested code.

\subsection{Area}
On a Virtex-7 XC7Z020, \tool consumes $4\%$ of the available registers and $6\%$ of available LUTs, which amounts to an average of 20\% additional logic overhead to the Pulpino SoC. $49$ 36Kbit Block RAM (BRAMs) are utilized, most of which are dedicated for the sparse loop path-indexed memories to ensure constant-time single-cycle access. Therefore, its width depends on the configured maximum number of indirect branches allowed in each loop path and number of bits required to encode them, as discussed in \sect{sect:indir_br}. In our implementation, the loop monitor is configured to tackle up to $4$ indirect branches and requires $10$ bits to encode them in $Path\_ID$, resulting in $16$ BRAMs per loop. Since we allow up to 3 levels of nested loops, we require $48$ BRAMs. Configuring these parameters to lower numbers reduces the memory requirements significantly at the expense of coarser granularity or additional logic overhead respectively. Alternatively, we are currently optimizing our implementation and leveraging content-addressable memories (CAMs) for these memories instead. This would remain to satisfy our requirement for constant-time access while also reducing the memory consumption significantly. However, implementing parallel CAM search is logic consuming and must be optimized such that it does not affect the maximum operating clock frequency of the entire architecture.




\subsection{Security}

The primary security requirement of \tool{} is to provide an \emph{accurate}, \emph{complete}, \emph{authentic}, and \emph{fresh} attestation of \prv{}'s control flow.
This requires an integrity-protected mechanism for recording control-flow information and unforgeably communicating this to \vrf{}.

\textbf{Control-Flow Recording.} 
One of the main contributions of \tool{} is using low-overhead hardware extensions to record control-flow information preventing it from being modified or subverted by malicious software. 
The on-chip memory employed by \tool{} for storing the \stp{} addresses prior to their hashing is also assumed to be protected from adversarial access. 
The hardware extensions are guaranteed to receive every control-flow event from the processor, thus ensuring that the complete control flow is recorded.  
All \stp{} addresses are cryptographically hashed resulting in the authenticator \auth{}. The auxiliary metadata \meta{} records (1)~the unique paths within each loop; (2)~the number of repetitions of each path; and (3)~all indirect branches encountered within loops. 

\textbf{Attestation Protocol.}
\tool{} makes use of the widely-used secure challenge-response attestation protocol.
As explained in~\sect{sect:model}, \prv{} sends the recorded program path $P$ along with a digital signature over $P$ and a nonce supplied by \vrf{}.
If \prv's signing key has not been compromised, this signature guarantees the authenticity of the attestation, and the inclusion of the challenge nonce ensures freshness.
Our assumed software adversary cannot compromise the signing key because it is stored in hardware-protected secure memory.
Any tampering with the attestation messages can be detected by \vrf{}.

Given that the control flow recording and the signing key is protected from software attacks, the resulting attestation report provided by \tool{} is accurate, complete, authentic, and fresh.
Since \prv{}'s code is immutable and is statically attested at boot time, \vrf{} has complete information about \prv{}'s execution.
As described in \sect{sect:model}, \vrf{} also has access to the CFG of the attested software, which it can use to identify permissible control flows and detect control-flow attacks or non-control data attacks.

%% file: 7_related.tex
\section{Related Work}\label{sect:related}


\textbf{Remote Attestation.} 
Most prior work focuses on \emph{static} remote attestation~\cite{ima,Eldefrawy12,Brasser2015}, which is orthogonal to run-time attestation -- the focus of this paper. 
Software-based attestation~\cite{Seshadri05a} can, under strict assumptions, enable static attestation of legacy devices without hardware-based trust anchors. 
Property-based attestation~\cite{Sadeghi04} can attest behavioral characteristics of a program, with the assistance of a trusted third-party. 
However, none of these can attest control-flow at machine code instruction level.


Prior work on run-time attestation focuses on specific aspects of a program's execution.
ReDAS~\cite{Kil09} attests program data invariants, such as the integrity of a function's base pointer, at each system call.
Trusted virtual containers~\cite{Bailey10} attest the run-time launch order of application modules -- a form of coarse-grained control-flow attestation that does not include internal control flows within modules.
DynIMA~\cite{Davi09} uses dynamic taint analysis and tracing to attest run-time properties that may be symptomatic of run-time attacks. However, it does not cover non-control data attacks and incurs high performance overhead due to dynamic taint analysis.


C-FLAT~\cite{Abera2016a} is a fine-grained control-flow attestation scheme. \tool\ also leverages the idea of attesting the control flow of an application by computing a cumulative hash of executed branches but with several fundamental differences. C-FLAT requires \emph{instrumentation} of all control-flow instructions thereby violating legacy compliance. In contrast, \tool\ does not require any binary rewriting. C-FLAT requires complete coverage in the offline binary analysis, as un-instrumented control-flow instructions could be exploited to mount undetectable attacks. This is not possible in \tool\ as every executed branch is monitored by design.
Finally, C-FLAT incurs significant performance overhead, whereas \tool incurs no performance overhead due to its efficient hardware support for control-flow attestation.

\textbf{Tracing and Debug Mechanisms.}
Intel processors provide the \emph{Last Branch Record} (LBR) and \emph{Branch Trace Store} (BTS) mechanisms, which can be used
to trace control-flow events~\cite{Soffa11}.
However, the overhead incurred by these debugging mechanisms makes them unsuitable for control-flow attestation.
Recently, Intel processors introduced \emph{Intel Processor Trace} (IPT)~\cite{Intel-IPT}, a low-overhead execution tracing feature that collects more tracing information than BTS (including execution mode and timing information).
However, IPT cannot be directly used for control-flow attestation as it only reports control-flow events that cannot be inferred from static analysis. ARM's CoreSight\footnote{https://www.arm.com/products/system-ip/coresight-debug-trace} debug and trace architecture provides a mechanism to access trace information from different hardware trace components. 
However, high-throughput tracing on ARM typically requires the use of proprietary hardware.

\textbf{Hardware-Assisted Security.}
Recent work~\cite{basakiccad2015,iips} developed a generic architecture for enforcing a diverse range of SoC security policies. 
Each IP block has an individually-customized security wrapper that sends security-relevant events and information to a central security controller to enforce individual security policies for each IP.
However, this incurs high memory and logic complexity overhead as the number of IPs increases.
It has further been proposed~\cite{basakdac2016,karri2015} that this could be made more practical by re-purposing design-for-debug features found on many SoCs -- a promising approach which could complement \tool{} in future.


Sofia~\cite{Sofia} is a recent hardware-assisted architecture for enforcing control-flow integrity (CFI).
It encrypts instructions with CFI-dependent data, such that they can only be decrypted at run-time as part of a valid control-flow path, and it ensures instruction integrity by checking MACs on groups of instructions at run-time.
However, unlike \tool{}, this requires software instrumentation and places decryption in the critical execution path, thus incurring total execution time overheads of up to 110\%. 

%% file: 8_conc.tex
\section{Conclusion}\label{sect:conc}
Due to the increasing prevalence of interconnected embedded systems, software running on these devices have become a prime target for remote attacks. We presented in this paper the first hardware-based control-flow attestation scheme that allows precise detection of remote memory corruption attacks in embedded system software. Our architecture, \tool, monitors, measures and reports the program's behavior by interfacing with the processor to intercept control-flow events. \tool\ does not require any code instrumentation (compliant to legacy software), compiler toolchain or instruction set extension. Our proof-of-concept implementation on the open-source RISC-V core is highly efficient with no performance impact on the attested software at the expense of minimal logic overhead and on-chip memory.\\
\textbf{Acknowledgments.} This work was supported by the German Science Foundation CRC 1119 CROSSING project, the German Federal Ministry of Education and Research (BMBF) within CRISP, the EU's Horizon 2020 research and innovation program under grant number 643964 (SUPERCLOUD), Tekes --- the Finnish Funding Agency for Innovation (CloSer project), and the Intel Collaborative Research Institute for Secure Computing (ICRI-SC).